\documentclass[aps,pra,twocolumn,floatfix]{revtex4-1}
\usepackage{amsmath,amssymb}
\usepackage{graphicx}
\usepackage{graphics}
\usepackage{epsfig}
\usepackage{color}
\usepackage{xcolor}
\usepackage{bm}
\usepackage[latin9]{inputenc}
\usepackage{ulem} % Crossing out with \sout{...}

\begin{document}

\title{Analysis of High-Contrast All-Optical Dual Wavelength Switching in Asymmetric Dual-Core Fibers}

\author{Le Xuan The Tai$^1$, Mattia Longobucco$^{2,3,4}$, Nguyen Viet Hung$^5$, Bartosz Paluba$^3$, Marek Trippenbach$^3$
, Boris A. Malomed$^{6, 7}$, Ignas Astrauskas$^8$, Audrius Pugzlys$^8$, Andrius Baltuska$^8$, Ryszard Buczynski$^{2,3}$, Ignac Bugar$^{8,9}$}

\affiliation{$^1$Faculty of Physics, Warsaw University of Technology, Koszykowa 75, 00-662 Warsaw, Poland.
\\
$^2$Department of Glass, Lukasiewicz Research Network Institute of Electronic Materials Technology, Wolczynska 133, 01-919 Warsaw, Poland.
\\
$^3$Faculty of Physics, University of Warsaw, Pasteura 5, 02-093 Warsaw, Poland.
\\
$^4$School of Electrical and Electronics Engineering, Nanyang Technological University, 50 Nanyang Avenue, 639798 Singapore.
\\
$^5$International Training Institute for Materials Science (ITIMS), Hanoi University of Science and Technology (HUST), No 1 - Dai Co Viet Str., Hanoi,Vietnam.
\\
$^6$Department of Physical Electronics, School of Electrical Engineering, Faculty of Engineering, and Center for Light-Matter Interaction, Tel Aviv University, Tel Aviv 69978, Israel.
\\
$^7$Instituto de Alta Investigacion, Universidad de Tarapaca, Casilla 7D, Arica, Chile.
\\
$^8$Photonics Institute, Vienna University of Technology, Gu$\beta$hausstra$\beta$e  25-29, Vienna, 1040, Austria.
\\
$^9$Institute of Chemistry and Environmental Sciences, University of Ss. Cyril and Methodius in Trnava, Nam. J. Herdu 2, 917 00 Trnava, Slovakia.}

\begin{abstract}
We systematically present experimental and theoretical results for the dual-wavelength switching of 1560 nm, 75 fs signal pulses (SPs) driven by 1030 nm, 270 fs control pulses (CPs) in a dual-core fiber (DCF). We demonstrate a switching contrast of 31.9 dB, corresponding to a propagation distance of 14 mm, achieved by launching temporally synchronized SP-CP pairs into the fast core of the DCF with moderate inter-core asymmetry. Our analysis employs a system of three coupled propagation equations to identify the compensation of the asymmetry by nonlinearity as the physical mechanism behind the efficient switching performance.
\end{abstract}

\maketitle
%\section{Introduction}
The realization of an all-optical switching in a simple fiber format has been a long-standing challenge in the field of nonlinear fiber optics \cite%
{wabnitzbook,li1,kieu,li2,Jensen1982}. The development of nonlinear directional couplers for all-optical switching holds great potential for many applications that require compact high-speed signal processing, without the use of free-space optics. Ultrafast nonlinear switching was proposed for femtosecond pulses in the normal-dispersion range of a step-index silica fiber coupler \cite%
{Friberg88}. The main limitations of this approach are a relatively high power and the intra-channel and inter-modal dispersion phenomena, which strongly affect pulses with widths $\sim 100$ fs. Various other devices have been developed, including metamaterial-based switches \cite{papaioannou}, ring resonators \cite{su}, plasmonic waveguides \cite{ono}, and nonlinear optical loop mirrors \cite%
{hirooka}. Nevertheless, when it comes to the fabrication simplicity and cascading, dual-core optical fibers
(DCFs) remain the most attractive option \cite{Trillo,Minardi2010,sarma,longobuccojlt}. In particular, a promising method of beam-by-beam cleaning, using orthogonally polarized fiber pulses, has been introduced
recently \cite{ferraro} applicable also for all-optical switching tasks. Our motivation for studying nonlinear DCFs is
stimulated by their potential for the reduction of the devices' length to
the scale of centimeters or millimeters, in contrast to the aforementioned
setups which require meter-scale fiber lengths \cite{hirooka,ferraro}.

In the experimental study \cite{longobuccojlt}, we had produced the first
evidence of dual-wavelength switching based on the interaction between two
temporarily synchronized pulses using an all-solid DCF. The operation mode
used a pair of femtosecond pulses simultaneously launched into the same
fiber core. A longer-wavelength ($1560$ nm) low-energy pulse served as a
signal pulse (SP), and the shorter-wavelength one ($1030$ nm) with higher
energy was a control pulse (CP). Using the synchronized pair of pulses with
the appropriate CP energy, SP switching between the excited and originally
idle cores was demonstrated, with negligible distortion. This scenario is
more efficient than the earlier reported straightforward energy control of a
single ultrashort pulse (\textit{self-switching}) \cite%
{Longobucco2019,longobuccooft,longobuccoao,tai1,tai2}, as it balances the
inter-core asymmetry. The asymmetry is inherent to DCFs, being a basic
limiting factor of the high-contrast directional-coupler performance.
Accordingly, the two cores of an asymmetric DCF are identified as fast and
slow ones, with effective refractive indices $n\textsubscript{f}<n%
\textsubscript{s}$. The motivation for
the use of the nonlinear dual-wavelength interaction is that, in the time window
defined by its duration, the
co-propagating CP reduces the group velocity of the fast core to the level of the slow one,
if the pulse pair is launched into the fast core \cite{longobuccojlt}.
%By this approach routing of the signal pulse is attainable alternating the OFF and ON state of the control pulse. In the OFF state the signal pulse remains dominantly in the excited core and in the ON state is switched to the non-excited core at fiber length equal to the coupling length at the signal wavelength.
An important aspect of this approach is the strong spectral dependence of
the coupling length, which prevents the energy transfer of the
shorter-wavelength CP pulses to the idle core. As a result, the switching
contrasts $>25$ dB was recorded, exceeding the best experimental results for
the self-switching in the C-band \cite{longobuccoao} and the theoretical
prediction for the ultrafast soliton self-trapping in highly nonlinear DCFs
\cite{Longobucco2019}. Additionally, shorter $14$ mm fibers were used,
instead of the $35$ mm one used in the study of self-switching at the signal
wavelength of $1550$ nm.

However, the dual-wavelength switching scheme has its drawbacks -- mainly,
relatively high CP energies, $\sim $ a few nJ. Here we present an advanced
experimental study, also supported by numerical simulations, in which the
fiber length, CP energy, and the CP-SP delay were simultaneously optimized,
thus leading to the reduction of the switching energy to the sub-nJ range,
while preserving the high switching-contrast level.

\begin{figure}[h]
\centering
\includegraphics[width=3.5in]{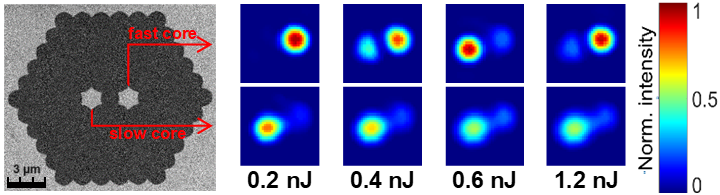}
\caption{Left: The scanning-electron-microscope image of the cross-section
of the all-solid DCF. Top and bottom series: IR-camera images of the
1560 nm, 75 fs SP field at the DCF output depending on energy of the 1030
nm, 270 fs CP exciting the fast and slow cores, respectively, of a 14 mm
length DCF.}
\label{fig1}
\end{figure}

In this work, we used simple-cladding all-solid DCFs with the
cross-section displayed in Fig. \ref{fig1} (left), similar to the samples
used in previous works \cite{longobuccojlt, longobuccoao}, which provide more details. The
distance between the core centers is 3.1 \textmu m, and the effective mode
area of both cores is $\approx 1.41$ \textmu m\textsuperscript{2}. The
fiber is made of two thermally matched soft glasses (lead-silicate and
borosilicate for the cores and cladding, respectively), with a very high
refractive-index contrast $\simeq 0.4$ between the core and cladding in the
C-band. The nonlinear refractive index in the core lead-silicate glass is $%
20 $ times higher than in the silica glass \cite{cimek}. The mismatch
between the fast and slow cores is $\delta n\textsubscript{a}=n%
\textsubscript{s}-n\textsubscript{f}=1.214\cdot 10^{-4}$. The experimental
setup includes two laser arms generating femtosecond CPs ($1030$ nm, $270$
fs ones produced by a commercial ultrafast Yb:KGW amplifier - Pharos, Light
Conversion) and SPs ($1560$ nm, $75$ fs ones from a self-made double-pass
optical parametric amplifier pumped by the second harmonic of the same
Yb:KGW amplifier), at the repetition rate of $10$ kHz. The two pulses were
combined by a dichroic mirror and synchronized by a delay line placed in the
control arm. The CP energy and polarization were managed independently,
using two half-wave plates and by a Glan-Taylor polarizer placed between
them. After passing the fiber, the control field was selectively filtered
out a by a high-reflectance mirror, securing that only the signal field from
the DCF output facet was imaged on a multimode collection fiber of a
spectrometer or infrared (IR) camera.

Similar to the findings in Ref. \cite{longobuccojlt}, a series of CP spatial and spectral distributions of the CP output field were separately recorded from both cores. In addition to investigating the effects of fiber length,  and the choice of the excited core (fast or slow), our study extends its focus
to examine the CP-SP delay effect. The camera images were
processed by calculating the dual-core extinction ratio ($\mathrm{ER}$),
separately integrating the intensity distribution in the area of each core.
Fig. \ref{fig1} (right) shows the camera images at the output facet of the DCF with
an optimal length of 14 mm, following the excitation of the fast and slow
cores (top and bottom series, respectively). The results reveal that the
excitation of the fast core provides efficient switching performance, thanks
to the above-mentioned DC asymmetry-balancing principle \cite{longobuccojlt}. The switching contrast (maximal $\Delta\mathrm{ER}$) is 31.9 dB between control-energy levels $0.2$ and $%
0.6$ nJ, at which the highest and lowest $\mathrm{ER}$\textrm{s} were
recorded, respectively. In the case of the slow core excitation (the bottom
series in Fig. 1), no switching performance was observed. Therefore, we
report results solely for the fast-core excitation.

%\section{Theoretical Insight}
%\label{sec:TM}

The DCF is modelled by coupled nonlinear Schr\"{o}dinger equations (NLSEs).
For CP amplitude $A_{0}$, which propagates only in the
excited core, NLSE is
\begin{equation}
\partial _{z}A_{0}=-\beta _{10}\partial _{t}A_{0}-(i/2)\beta _{20}\partial
_{t}^{2}A_{0}+i\gamma |A_{0}|^{2}A_{0},  \label{eq:control_phys}
\end{equation}%
where $z$ and $t$ are the propagation distance and time in physical units,
and coefficients $\beta _{10}$, and $\beta _{20}$, $\gamma $, represent the
inverse group velocity, group-velocity dispersion (GVD), and SPM nonlinearity, respectively. NLSEs for the SP amplitudes in the excited and idle cores (in the simulation framework are always the fast and slow, respectively) are
\begin{gather}
\partial _{z}A_{f}=-\beta _{1f}\partial _{t}A_{f}-(i/2)\beta _{2f}\partial
_{t}^{2}A_{f}+  \notag \\
i\kappa _{0}A_{s}-\kappa _{1}\partial _{t}A_{s}-2i\delta A_{f}+i\gamma
|A_{0}|^{2}A_{f},  \notag \\
\partial _{z}A_{s}=-\beta _{1s}\partial _{t}A_{s}-(i/2)\beta _{2s}\partial
_{t}^{2}A_{s}+  \notag \\
i\kappa _{0}A_{f}-\kappa _{1}\partial _{t}A_{f},  \label{eq:signal_phys}
\end{gather}%
where $\beta _{1f}=\beta _{1s}\equiv \beta _{1}\neq \beta _{10}$ is the SP
inverse group velocity, $\beta _{2f}=\beta _{2s}=\beta _{2}$ is the SP\ GVD,
$\gamma $ is the XPM coefficient, $\delta =(\beta _{0f}-\beta _{0s})/2$ is
the inter-core propagation constant difference, $\kappa _{0}$ and $\kappa _{1}$
being zeroth- and first-order coupling coefficients.

By means of rescaling and introducing retarded time $\tau=$ $\sqrt{\kappa
_{0}/|\beta _{2}|}t-\beta _{10}z$, and rescaling $\zeta=\kappa _{0}z$, $\psi =\sqrt{%
\gamma /\kappa _{0}}A_{0}$ and $\phi _{f,s}=\sqrt{\gamma /\kappa _{0}}%
A_{f,s} $, Eqs. (\ref{eq:control_phys}) and (\ref{eq:signal_phys}) are cast
in the normalized form, with $\beta _{2}=\gamma =\kappa \equiv 1$. In
particular, the rescaled equation for CP is

\begin{equation}
	i\partial _{\zeta}\psi =\frac{\beta
	_{20}}{2\beta _{2}}\partial _{\tau}^{2}\psi -|\psi |^{2}\psi. 
\end{equation}

The CP input in the fast core, with amplitude $a_{CP}$ and FWHM temporal size
$T_{CP}$ is $\psi (z=0,\tau )=a_{CP}\exp [-(\tau /w_{CP})^{2}]$, where $%
w_{CP}=T_{CP}/\left( 1.1774\sqrt{|\beta _{2}|/\kappa _{0}}\right) $. The set
of the rescaled equations contains three adjustable parameters: the
inverse-group-velocity CP-SP mismatch $\alpha =(\beta _{1}-\beta _{10})/%
\sqrt{|\beta |/\kappa _{0}}$, the inter-core index mismatch $\sigma =\delta
/\kappa _{0}$, and the dispersive coupling coefficient $\epsilon =\kappa
_{1}/\sqrt{\kappa _{0}|\beta _{2}|}$.

The SP\ input with amplitude $a_{SP}$ and FWHM size $T_{SP}$ in the fast
channel is $\phi _{f}(0,\tau )=a_{SP}\exp \{-[(\tau -d)/w_{SP}]^{2}\}$, where $%
w_{SP}=T_{SP}/\left( 1.1774\sqrt{|\beta _{2}|/\kappa _{0}}\right) $ and $d$ is
the CP-SP delay. The units of the propagation length and time in the
experiment are $t_{0}\equiv \sqrt{|\beta _{2}|/\kappa _{0}}=24.9854$ fs and $%
z_{0}\equiv 1/\kappa _{0}=8.11402$ mm. The corresponding CP\ energy is

\begin{eqnarray}
	E=\int_{-\infty }^{+\infty }|A_{0}(0,\tau)|^{2}t_{0}d\tau &=&\kappa _{0}\tau _{0}a_{CP}^{2}w_{CP}\gamma ^{-1} \notag \\
                     &\approx& 15.189a_{CP}^{2}\;\text{pJ}.
\end{eqnarray}

The CP and SP widths are $T_{CP}=270$ fs and $T_{SP}=75$ fs. The effective coupling is $\kappa _{\mathrm{eff}}\equiv \sqrt{\kappa _{0}^{2}+\delta ^{2}} =0.90098$ mm$^{-1}$, where $\kappa _{0}=0.12324$ mm$^{-1}$. In the scaled units, the widths are $w_{CP}=9.1780$, $%
w_{SP}=2.5494$.

\begin{figure}[h]
\centering
\includegraphics[width=3.5in]{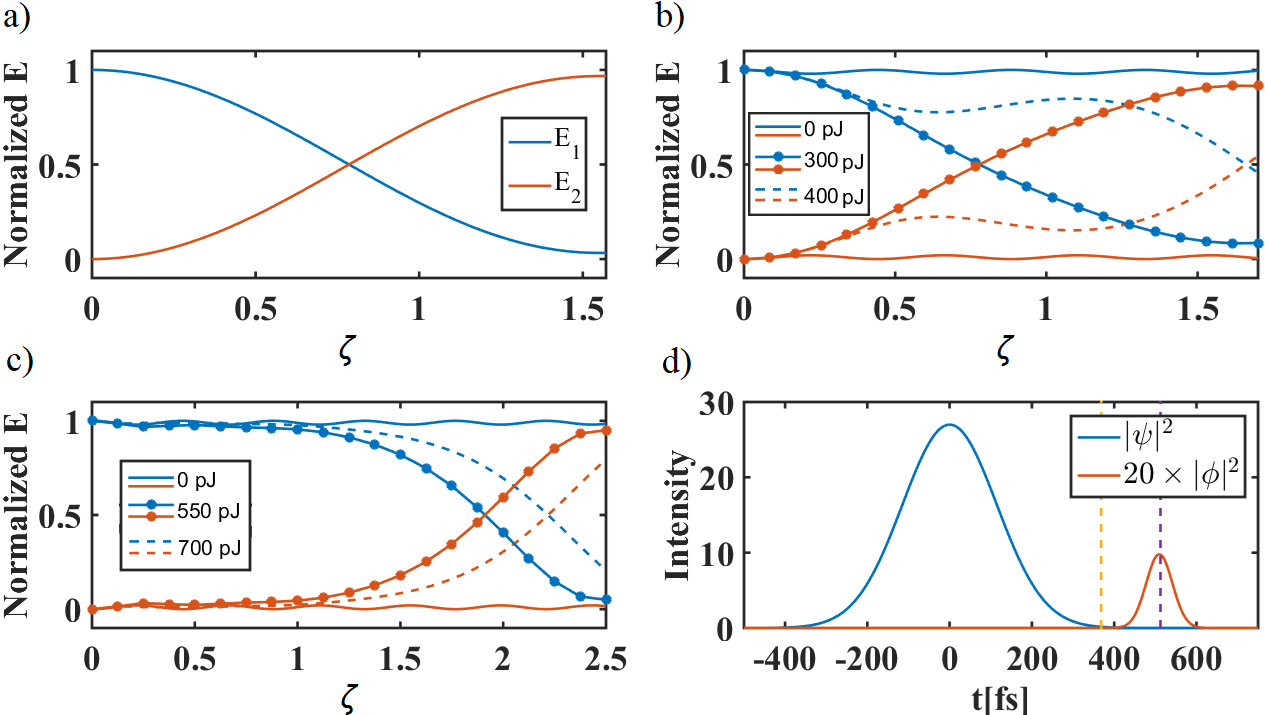}
\caption{Simulated switching performance in the DCF: a) Symmetric coupler: perfect cross-transfer of the SP. b) Asymmetric coupler: oscillations with negligible transfer in the absence of CP (continuous lines), the best performance provided by CP
energy of $300$ pJ (dotted lines). Dashed lines illustrate the situation when the CP energy is too high (400pJ). c) Compensation of the SP mismatch by the CP of 550 pJ energy for longer DCFs. d) CP and SP shapes. The distance between vertical dashed lines displays their relative walk-off over the DCF length.}
\label{fig2}
\end{figure}

We simulated the pulse propagation in the parameter range corresponding to
the experiment performed with the $14$ mm long fiber and nonlinearity
coefficient $\gamma =1.86\,(\mbox{W}\cdot \mbox{m})^{-1}$. Parameter values
for the simulations were produced by Lumerical at the CP and SP wavelengths $%
\lambda =1030$ nm and $1560$ nm, respectively. While CP is affected by GVD
and SPM, for the relatively weak SP, the only essential nonlinear effect is
the XPM interaction with CP in the fast core. Results of the simulations are
presented in Fig. {\ref{fig2}}. Panel a) reports cross-transfer of the SP in
the symmetric DCF in the course of half a period of the inter-core
oscillations. In the asymmetric DCF, where the oscillation period decreases
due to mismatch, the transfer is dramatically reduced, as shown by the
continuous wavy lines\ in panel b). The CP with appropriate energy
compensates the mismatch in the asymmetric DCF, as shown by the dotted lines
in the same panel. If the CP energy is not optimized, the switching
performance may still be poor, as shown by the dashed lines. The
optimization has to be performed by adjusting the CP-SP delay, fiber length,
and CP energy. Panel c) shows that the optimization is also possible for
longer DCFs. In this case, more oscillations are observed in\ the absence of
CP, and effective switching is achieved with an appropriate CP energy. In
the last panel of Fig. \ref{fig2}, we display the CP and SP shapes and their
relative walkoff, when they propagate together along the fiber, for the
parameters corresponding to our experiment.

In the absence of the CP, SPM, and higher-order coupling, the coupled
equations admits an exact solution, which includes both the mismatch and
dispersion. It is given by a two-component chirped Gaussian pulse moving
along the temporal coordinate, with the components periodically oscillating
between the cores:

\begin{align}
\phi_{1}=& \Phi (\zeta)\left[ i\sigma \sin \left( \sqrt{1+\sigma ^{2}}\zeta\right) +%
\sqrt{1+\sigma ^{2}}\cos \left( \sqrt{1+\sigma ^{2}}\zeta\right) \right] ,
\notag \\
\phi_{2}=& -i\Phi (\zeta)\sin \left( \sqrt{1+\sigma ^{2}}\zeta\right) ,  \label{Psi}
\end{align}%
where $\Phi (\zeta )=\frac{W}{\sqrt{W^{2}+i\zeta }}\exp [-\frac{(T+\Omega
\zeta )^{2}}{2(W^{2}+i\zeta )}-i\Omega T-\frac{i}{2}\Omega ^{2}\zeta ]$.
This solution depends on two arbitrary parameters, width $W$ and frequency
shift $\Omega $. The oscillation period is clearly mismatch ($\sigma $--)
dependent, cf. Fig. \ref{fig2}.

%-------------
%-------------

\begin{figure}[h]
\centering
\includegraphics[width=3.5in]{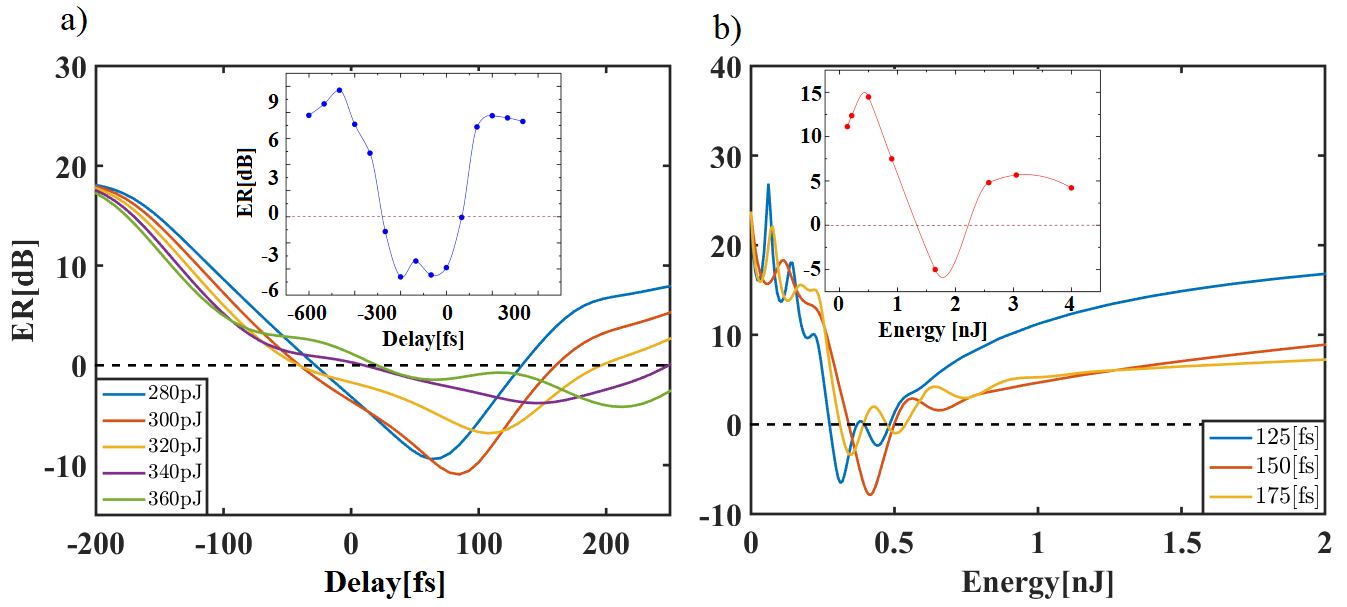}
\caption{Simulated $\mathrm{ER}$ dependence in the $14$ mm	long DCF  on: a) the CP-SP delay for different CP energies; b) the CP energy for different delays. 
Insets present the experimental results for the CP energy $490$ pJ in a), and for delay of  167 fs in
b).}
\label{fig3}
\end{figure}

To further characterize the
switching performance, we address the dependence of the output extinction ratio,
$\mathrm{ER}=10\log (E_{f}/E_{s})$, on the CP-SP delay and CP power. The
theoretical and experimental dependences on\ the delay are reported in Fig. %
\ref{fig3}a), showing that the $\mathrm{ER}$ is minimized at larger values
of the delay with increasing the CP energy, in agreement with the
above-mentioned asymmetry-compensation principle. Indeed, at higher energies,
the SP experiences the same refractive-index change at a larger delay in the
CP tail. The simulations predict a minimum of $\mathrm{ER}$ at the level of $%
-10$ dB at the energies in the range of $280-300$ pJ and around the delay of
$75$ fs. Taking into regard the $150$ fs walkoff produced by the $14$ mm
long DCF \cite{longobuccojlt}, such conditions imply the SP moving in the
peak area of the CP. Experimental results do not reveal so low $\mathrm{ER}$
because of additional CP nonlinear distortions (not included in the model),
that may be significant around the CP peak. On the other hand, the predicted
delay dependence at $320$ pJ resembles the experimental curve (the inset in
Fig. \ref{fig3}a) both in terms of the $\mathrm{ER}$ minimum level and delay
range where $\mathrm{ER}$ is negative. Note that the experimental curve was
obtained for $490$ pJ CP energy, which also indicates that distortions
affect the CP, impeding to reach the compensation effect at higher
energies. In Fig. \ref{fig3}b) we plot the $\mathrm{ER}$ dependence on the
CP energy for the CP-SP delay in the range of $125-175$ fs. These curves
confirm that the $\mathrm{ER}$ minima are experimentally accessible with the
$150$ fs delay, which is an essential result, as the absolute delay was not
identified in the experiment. Furthermore, getting the best experimental
result at the large delay suggests a possibility of improving our approach by
suppressing the walkoff. It is evident that, at zero walkoff, the two pulses
should be launched without delay. This condition would allow the use of
significantly lower CP energies, helping to mitigate the nonlinear
distortions. A general inference is that there is
a wide range of both CP-SP delays and CP energies where the high-contrast
(low-$\mathrm{ER}$) switching occurs in the robust form, i.e., when slight
changes of these parameters do not compromise it.

%-------------
%-------------

\begin{figure}[h]
\centering
\includegraphics[width=3.5in]{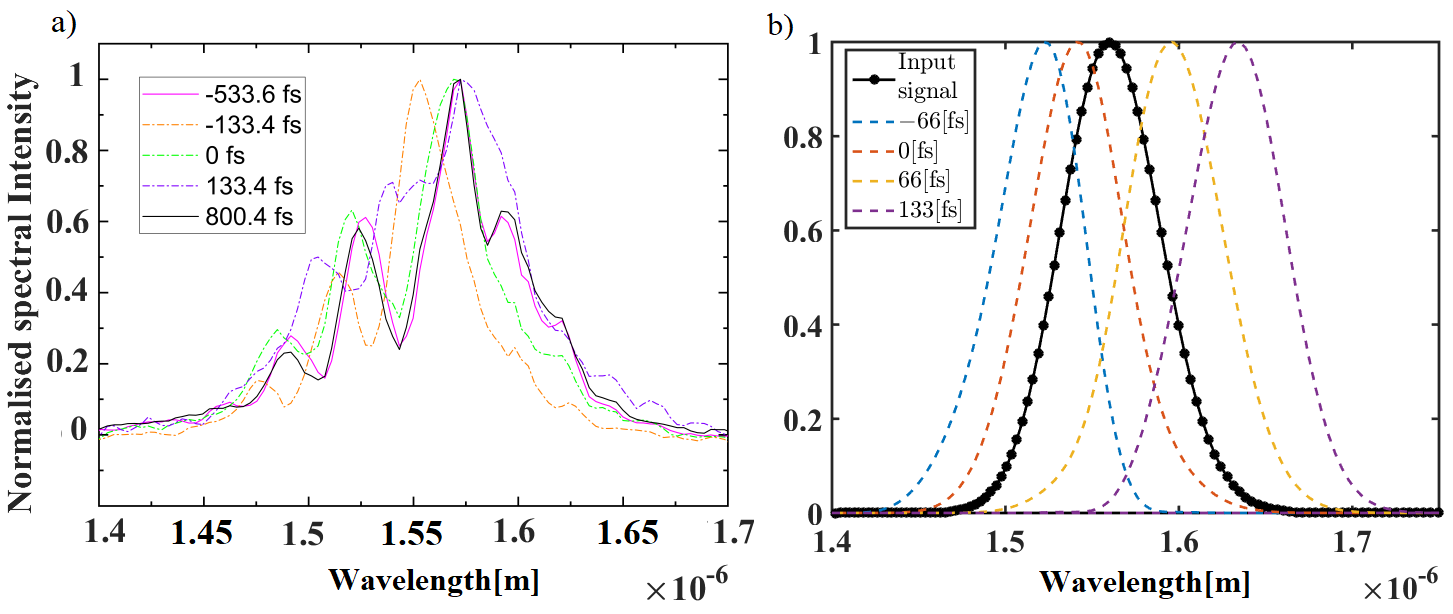}
\caption{Normalized SP\ spectra for different CP-SP delays at the output of
the originally excited and idle cores (solid and dashed lines, respectively)
at $14$ mm DCF length. a) Experimental results at the CP
energy of $600$ pJ. b) Simulations for the CP energy of $320$ pJ. Other
parameters are $\protect\alpha =-3.5$, $\protect\epsilon =0.3$ and $\protect%
\sigma =7.033$ corresponding to experimental values $\protect\beta
\textsubscript{1}-\protect\beta \textsubscript{10}=-111$ ps/m, $\protect%
\kappa _{1}=-0.921$ ps/m, $\protect\delta =865$ m$^{-1}$.}
\label{fig4}
\end{figure}

Finally, we have performed simulations varying the delay for $\sigma =7.033 $%
, $\alpha =-3.5$, $E\textsubscript{CP}=320$ pJ, at $14$ mm DCF length, and
analyzed the spectral shape of the output SPs. The results are reported in
Fig. \ref{fig4}b, where the experimental spectra are in panel a). The
numerical results, which are in qualitative agreement with the experimental
counterparts, produce, for positive delays, a red-shifted and slightly
broader SP\ spectrum. In the case of the negative delay, also the predicted
blue shift and spectral narrowing were observed. It is worth mentioning that the solid lines are nearly identical in the case of experimental spectra and were registered at delays -533.6 fs and 800.4 fs from the excited core. Thus, they represent situations without overlap between CP-SP pulses and can be considered as original SP spectra.
%-------------
%-------------

Thus, we have conducted a detailed analysis of the dual-wavelength switching
of $1560$ nm, $75$ fs SPs in the DCF, using the interaction with $1030$ nm, $%
270$ fs CPs as a nonlinear drive. Effects of the fiber length, CP energy, and CP-SP\
time delay on the switching performance have been studied, experimentally
and numerically. The highest switching contrast of $31.9$ dB was achieved in
the $14$ mm long DCF, with a broadband character in the spectral range of $%
1450-1650$ nm. The simulations reveal the role of the inter-core difference
in the effective refractive index and CP-SP walk-off, confirming that the
efficient  switching is provided by the nonlinear balance of the inter-core
asymmetry. Furthermore, the simultaneous effect of the CP
energy and CP-SP delay on the dual-core $\mathrm{ER}$ was identified. One of the key
advantages of our approach is the moderate nonlinear CP-SP\ interaction,
which only slightly deforms the signal field. The numerical results support
this concept, producing moderate changes in the SP spectra. On the other hand, the experiment
indicates more complex nonlinear distortions of CPs than predicted by the
model and roughly its two times higher energy level. The reason of this discrepancy is that the presented model does not take into account the linear and nonlinear dissipative effects \cite{tai2} in order to preserve the numerical stability of the rather complex three wave interaction in two coupled channels. Nevertheless, in the region of moderate nonlinear interaction -
namely, when the SP overlaps with a CP's tail (delays exceeding 150 fs, Fig. \ref{fig3}) - the agreement between the
experimental and numerical results is convincing.

To summarize, our study provides a comprehensive analysis of the
dual-wavelength switching in the appropriately designed DCF. Our
experimental and theoretical findings shed light on the physical mechanisms
behind the observed phenomenology and highlight the advantages of the proposed
approach, which makes it possible to achieve efficient switching while
avoiding conspicuous distortion of the signal field. In the framework of the
reported experiments, better matching of the SP and CP beams was ensured resulting in equal beam diameters at the DCF input. As a
result, the energies at which the switching takes place are $\simeq 10$
times lower than those reported previously \cite{longobuccojlt}. The sub-nJ,
high-switching-contrast findings presented in this paper offer the
application to the design of an all-optical signal-processing scheme.
It may support a processing rate above 2 THz/s, primarily limited by the duration of the CP. By eliminating the walkoff, CP pulse durations below 100 fs can be applied, supporting processing rates up to 10 THz/s. Therefore, our approach offers significant advancements, primarily for the time-division-multiplexing tasks. However, the application to the
wavelength-division multiplexing is limited by the femtosecond pulse duration and nonlinear interactions, resulting in pulse bandwidths
$\sim 100$ nm. The theory presented here predicts the possibility of further improvements in the switching contrast, while simultaneously reducing the CP energy, by suppressing the walkoff between the control and signal pulses.
However, achieving such improvements will necessitate a meticulous
engineering of the DCF design, optimizing both its dispersion and coupling
characteristics. The development of such a new generation of DCFs may offer
new applications to ultrafast signal processing and time-resolved
spectroscopy.
\newline
\newline
\textbf{Funding} Israel Science Foundation (1695/22 - B.A.M.). Polish National Science Center (2019/33/N/ST7/03142 - M.L. - 2020/02/Y/ST7/00136 - R.B.). Vietnam Ministry of Education and Training (MOET) (B2022-BKA-14 - N.V.H.). Austrian Science Fund (FWF) (I 5453-N - I.B., A.P., A.B.). University of Warsaw Integrated Development Programme (ZIP) - 2nd round competition for doctoral students exchange travels (M.L.). Slovak Scientific Grant Agency (VEGA 2/0070/21 - I.B.) \newline
\newline
\textbf{Disclosures} The authors declare no conflicts of interest. \newline
\newline
\textbf{Data Availability} Data underlying the results presented in this paper are not publicly available at this time but may be obtained from the authors upon reasonable request.

% Bibliography

\end{document}